# A Normalized Sunspot-Area Series Starting in 1832: an Update

V.M.S. Carrasco[1,2] • J.M. Vaquero[2,3] • M.C. Gallego[1,2] • F.Sánchez-Bajo[4]

[1] Departamento de Física, Universidad de Extremadura, Badajoz, Spain

[2] Instituto Universitario de Investigación del Agua, Cambio Climático y Sostenibilidad (IACYS), Universidad de Extremadura, Badajoz, Spain

[3] Departamento de Física, Universidad de Extremadura, Mérida (Badajoz), Spain (jvaquero@unex.es)

[4] Departamento de Física Aplicada, Universidad de Extremadura, Badajoz, Spain

Abstract. A new normalized sunspot-area series has been reconstructed from the series obtained by the Royal Greenwich Observatory and other contemporary institutions for the period 1874‐2008 and the area series compiled by De la Rue, Stewart, and Loewy from 1832 to 1868. Since the two sets of series do not overlap in time, we used as a link between them the new version of sunspot index number (Version 2) published by SILSO (Sunspot Index and Long-term Solar Observations). We also present a spectral analysis of the normalized area series in search of periodicities beyond the well-known solar cycle of 11 years and a study of the Waldmeier effect in the new version of sunspot-number and the sunspot-area series presented in this study. We conclude that while this effect is significant in the new series of sunspot number, it has a weak relationship with the sunspot-area series.

## 1. Introduction

The historical reconstruction of solar activity is a key element to understand the Sun's behavior and effects on Earth. Moreover, the study of the past behavior of our Sun is evidently very important from the point of view of the modern astrophysics (Vaquero and Vázquez, 2009; Usoskin, 2013). In this sense, there are several series that provide information about solar activity of the past from direct observation of sunspots. For example, the record of naked-eye sunspots observations, although they are not abundant, covers the last two millennia (Clark and Stephenson, 1978; Vaquero, Gallego, and García, 2002). However, it was not until the invention of the telescope, about 400 years ago, that systematic observations of the Sun began to be carried out (Hoyt and Schatten, 1998; Vaquero, 2007).

Considering the sunspot counts to establish a value representing the state of solar activity, the commonly used indices are the international sunspot number SSN and the group sunspot number (GSN) (Hoyt and Schatten, 1998; Clette *et al*., 2014). These indices are defined by $SSN = k(10g + s)$ and $GSN = k'12.08g$, where *s* represents the number of individual spots, *g* is the number of groups, and *k, k'* are correction factors that depend on the observer, telescope, seeing, etc. However, due to the evident discrepancies between the two series in the historical period, there has recently been carried out a review of both series with the aim of reconciling these indices. After this revision, Clette *et al*. (2015) showed that the two new series of solar activity, sunspot number [$S_N$], and group number [$G_N$], have a behavior closer than their predecessors. Furthermore, in this way, Usoskin *et al*. (2016) have presented a new sunspot-group series from statistics of active-day fractions.

Occasionally, data with respect to the area occupied by the spots on the solar disk can also be retrieved from historical sources (Arlt *et al*., 2013; Aparicio *et al*., 2014; Carrasco *et al*., 2014; Lefèvre and Clette, 2014). The records of sunspot areas are valuable because they have utility in different studies of solar physics, for example, the solar irradiance (Krivova, Balmaceda, and Solanki, 2007). Balmaceda *et al*. (2009) compared measurements of sunspot areas made at different observatories around the world to obtain a homogeneous single series. The series presented by Balmaceda *et al*. (2009) starts in 1874, when the Royal Greenwich Observatory initiated this kind of regular observations, and ends in 2008. However, systematic records of sunspot areas exist prior to 1874 (Casas and Vaquero, 2014), which can be considered to extend the series proposed by Balmaceda *et al*. (2009).

Vaquero, Gallego, and Sánchez-Bajo (2004) proposed a method to obtain a homogeneous sunspot-area series since 1832, using the available information about solar activity in the past. However, this study can be improved nowadays using the new results about sunspot area (Balmaceda *et al*., 2009) and sunspot number (Clette *et al*., 2015). The main objective of this study is extending the sunspot-area series of Balmaceda *et al*. (2009) from sunspot-area measures compiled by De la Rue, Stewart, and Loewy (1870) for the period 1832 – 1868. Moreover, we analyze whether the Waldmeier effect is present in sunspot-area series as has been found, for example, in the sunspot number (Waldmeier, 1955). In Section 2, we describe the data used in this study. Section 3 is devoted to the analysis of normalization of data and the presentation of the final series of sunspot areas. In this section, we also show a spectral analysis to find periodicities in the final series normalized, and we carefully study the Waldmeier effect in area series. Finally, Section 4 contains the main conclusions of this study.

**2. Data**

The Royal Greenwich Observatory conducted a program of systematic sunspot observations during the period 1874 – 1976 (Erwin *et al*., 2013; Willis *et al*., 2013a; 2013b). Among the parameters recorded at this observatory, we can find measurements of the area covered by sunspots on the solar disk. In addition to the area records from Royal Greenwich Observatory, Balmaceda *et al*. (2009) used data from other observatories (for example, Rome, Yunnan, or Catania) to construct a homogeneous series of sunspot areas. To homogenize all records, Balmaceda *et al*. (2009) carried out an intercomparison between the data of the selected observatories using as central the measurements made at the Royal Greenwich Observatory. Thus, they obtained a homogeneous series of daily sunspot areas with high temporal coverage for the period 1874–2008.

De la Rue, Stewart, and Loewy (1870) collected fortnightly data on sunspot areas for the period 1832–1868. This period is divided into three steps that comprise observations made from Schwabe (1832 – 1853), Carrington (1854 – 1860), and Kew Observatory (1861 – 1868). The reliability of these observations was analyzed by Vaquero, Sánchez-Bajo, and Gallego (2002).

In this article, we have combined the sunspot area series compiled by De la Rue, Stewart, and Loewy (1870) and Balmaceda *et al*. (2009) in order to obtain a homogeneous monthly series of sunspot areas that covers approximately the last two centuries. These series are not temporally overlapped because they cover the periods 1832–1868 and 1874–2008, respectively. However, to accomplish this task, we use as a link between the two series the new series of sunspot number (Clette *et al.*, 2015).

## 3. Analysis and Results

### 3.1. Mathematical Procedure

A direct comparison between the series of compiled sunspot areas by De la Rue, Stewart, and Loewy (1870) and Balmaceda *et al.* (2009) is not possible since, as we mentioned above, they do not overlap in time. However, we can set as a link between the two series the sunspot-number index due to its high correlation ($r = 0.97$) with the sunspot-area series. Recently, there has been conducted a review of the sunspot number indices of which the final result has been the publication of new versions of the latter (Clette *et al.*, 2015; Svalgaard and Schatten, 2016; Usoskin *et al.*, 2016). In this study, we have used the new version of the sunspot number index ($S_N$, Version 2) available on the web (www.sidc.be/silso/datafiles).

To carry out the normalization, we assume a potential dependence, on the one hand, between the Balmaceda area series [$SSA$] and the sunspot number index ($S_N$) for the period 1874–2008 such that $SSA = k \cdot (S_N)^C$, and, on the other hand, between the De la Rue area series [$SSA_R$] and sunspot number for each individual stage in the De la Rue, Stewart, and Loewy (1870) compilation: $S_{Ni} = a_i (SSA_{Ri})^{bi}$, where $i = 1$ to 3 (1 for 1832–1853, 2 for 1854–1860, and 3 for 1861–1868). Thereby, the normalized sunspot area series is defined by $SSA$ for the period 1874–2008, $SSA_N = k (S_N)^C$ for the period 1868–1874, and $SSA_N = k (a_i (SSA_{Ri})^{bi})^C$ for the period 1832–1868. The coefficients $k$, $c$, $a_i$, and $b_i$ are determined by potential fits for each sunspot number series.

### 3.2. Normalized Sunspot-Area Series

Thus, the expression for the best potential fit by comparing the monthly values of $SSA$ with $S_N$ for the period 1874–2008 is: $SSA = (2.5 \pm 1.0) S_N^{(1.29 \pm 0.01)}$, $r = 0.971$, $p$-value < 0.001. On the other hand, the best potential fits between the monthly values of $S_N$ and $SSA_R$ for each individual stage for the period 1832–1868 are: i) $S_{N1} = (0.4 \pm 1.1)$ $SSA_{R1}^{(0.86 \pm 0.02)}$, $r = 0.944$, $p$-value $< 0.001$; ii) $S_{N2} = (2.5 \pm 1.2)\, SSA_{R2}^{(0.57 \pm 0.03)}$, $r = 0.925$, $p$-value $< 0.001$; iii) $S_{N3} = (0.5 \pm 1.3)\, SSA_{R3}^{(0.76 \pm 0.04)}$, $r = 0.881$, $p$-value $< 0.001$. The correlation coefficient values found here are higher than those values obtained by Vaquero, Gallego, and Sánchez-Bajo (2004) using the old versions of the sunspot-number and area series and assuming a linear dependence between both indices. According to these results and following the procedure explained above, we have constructed a normalized monthly sunspot area series for the period 1832–2008 (Figure 1). The annual values for sunspot-area series normalized during the period 1832–1873 are presented in Table 1. Note that the units of areas are millionths of solar hemisphere (msh).

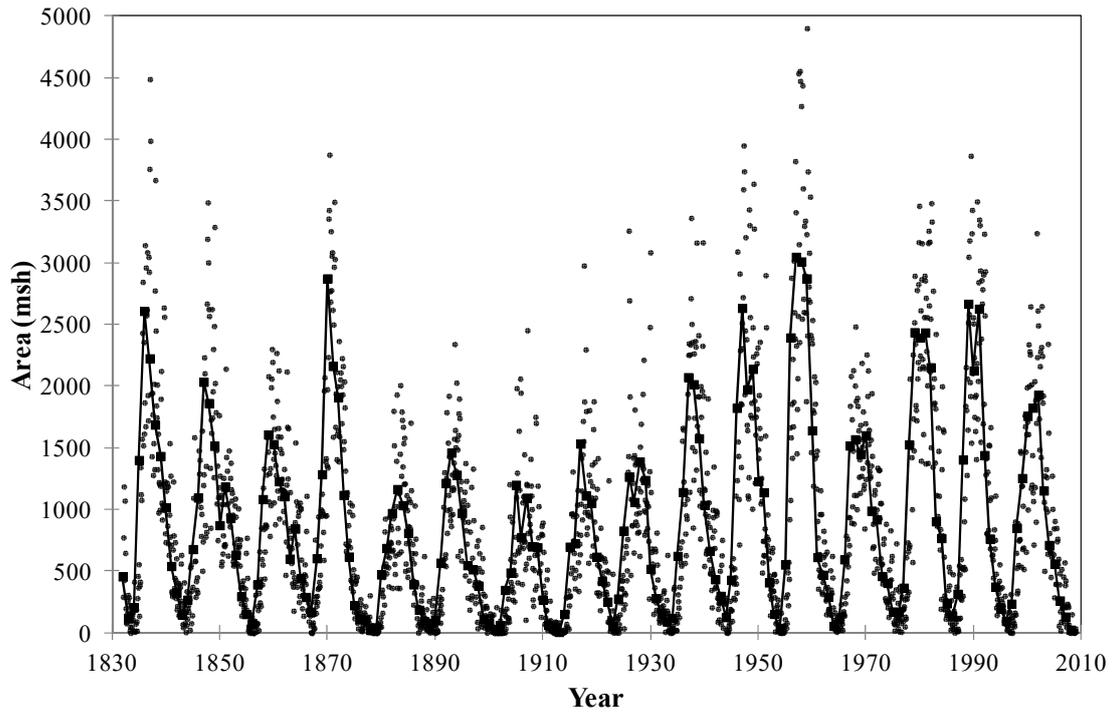

Figure 1. Sunspot area series normalized for the period 1832–2008. Gray dots represent the monthly values and black squares the annual values.

Table 1. Annual values for sunspot area series normalized during the period 1832 – 1873.

| Year | Area | Year | Area | Year | Area | Year | Area |
|------|------|------|------|------|------|------|------|
| 1832 | 457  | 1843 | 148  | 1854 | 296  | 1865 | 441  |
| 1833 | 121  | 1844 | 269  | 1855 | 153  | 1866 | 291  |
| 1834 | 208  | 1845 | 679  | 1856 | 69   | 1867 | 171  |
| 1835 | 1402 | 1846 | 1098 | 1857 | 392  | 1868 | 605  |
| 1836 | 2609 | 1847 | 2035 | 1858 | 1084 | 1869 | 1286 |
| 1837 | 2226 | 1848 | 1863 | 1859 | 1609 | 1870 | 2874 |
| 1838 | 1688 | 1849 | 1516 | 1860 | 1531 | 1871 | 2163 |
| 1839 | 1434 | 1850 | 873  | 1861 | 1229 | 1872 | 1912 |
| 1840 | 1018 | 1851 | 1188 | 1862 | 1107 | 1873 | 1117 |
| 1841 | 542  | 1852 | 932  | 1863 | 602  |      |      |
| 1842 | 326  | 1853 | 634  | 1864 | 845  |      |      |

## 3.3. Spectral Analysis

The periodicities in the solar activity have been known for a long time. Even periodic signals shorter than the known long-term variabilities as Schwabe (throughout 11 years) or Gleissberg (throughout 80 years) have been presented in several studies (Hathaway, 2015). For example, Kilcik *et al*. (2014) analyzed sunspot counts classified in several categories and found a 300-day periodicity in the counts of large and well-developed sunspot groups. Rieger *et al*. (1984) found a periodicity of 154 days (Rieger periodicity) after analyzing 139 gamma-ray solar flares seen from the *Solar Maximum Missions* (SMM) spacecraft and, later, Bai and Cliver (1990) also detected this signal from proton flares. Furthermore, Carbonell and Ballester (1992) analyzed the historical records of sunspot areas belonging to Solar Cycles 12 – 21 made at the Royal Greenwich Observatory in order to search for the periodicity of throughout 150 days in other solar-activity indices. That study revealed that this periodicity seems relevant for the period corresponding to Solar Cycles 16 – 21, however, it was not detected for Solar Cycles 12 – 15. For previous solar cycles, Ballester, Oliver, and Baudin (1999) discovered this periodicity Solar Cycles 2 using the Group Sunspot Numbers and Vaquero *et al*. (2010) found this periodicity in Solar Cycles 3 and 4 using historical observations of aurorae.

We have carried out a spectral analysis using the Morlet wavelet-analysis method (Torrence and Compo, 1998) to study the periodicities of the sunspot-area series presented in this study for the period 1832 – 1874. Figure 2 shows clearly with a significance level of 95 % the 11-year solar cycle and also a period of about 300 days around the maximum of Solar Cycles 8 and 9 that could be considered a harmonic of the periodicity around 150 days. In addition, other weak signals with a period about 100 days are detected in the maximum of the cycles. These results are very similar to those achieved by Vaquero, Gallego, and Sánchez-Bajo (2004).

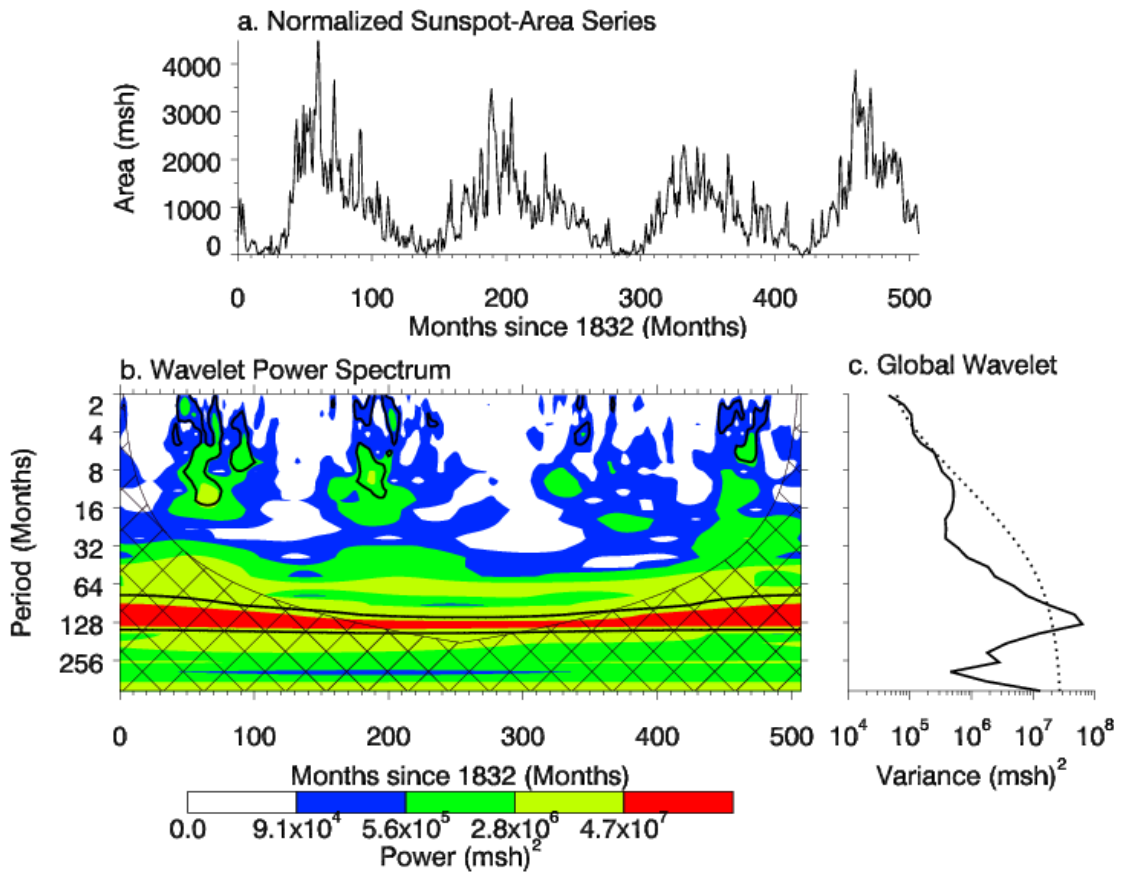

Figure 2. Morlet wavelet analysis for the sunspot-area series normalized during the period 1832 – 1874. We show (a) the sunspot area-series normalized, (b) wavelet power spectrum and (c) global wavelet spectrum. The contour levels are chosen so that 75 %, 50 %, 25 %, and 5 % of the wavelet power is above each level, respectively. Black contour is the 5 % significance level, using a red-noise background spectrum. The cross-hatched region is the cone of influence.

## 3.4 Waldmeier Effect

The inverse correlation between the rise time and the maximum amplitude of a cycle is known as the Waldmeier effect (Waldmeier, 1955). This effect involves the observation that the cycles with higher peaks of maximum solar activity have shorter rise times and, weaker cycles have longer rise times. Karak and Choudhuri (2011) provided a theoretical explanation of this effect using a dynamo model. The Waldmeier effect has been observed in the sunspot-number index (Kane, 2008) but not, for example, in the sunspot-area series (Dikpati, Gilman, and de Toma, 2008). There are also some studies using this effect to predict the maximum amplitude of solar cycles at the beginning of each cycle (Wilson, Hathaway, and Reichmann, 1998) and, therefore, it is necessary to know whether the Waldmeier effect works precisely. In this study, we have conducted a review of this phenomenon through the calculation of the parameters involved in this effect from the recently published series of the sunspot number index (Clette *et al.*, 2015) and the normalized sunspot area series proposed in this article.

First, to establish both the rise time and maximum amplitude of solar activity of each cycle, we have smoothed both series. The most widely used smoothing method is the 13-month running mean that assigns weights equal to one for the months from -5 to +5 and half weight for the months -6 and +6. However, this smoothing does not work well for high-frequency variations. In this sense, the Gaussian filters are preferable because they reduce such variations. Hathaway (2015) shows a tapered Gaussian filter given by:

$$W(t) = e^{-t^2/2a^2} - e^{-2}(3 - t^2/2a^2)$$

with $-2a + 1 \leq t \leq +2a - 1$, where *t* is the time from the center of the filter and 2*a* is the full width at half maximum (FWHM) of the filter. The significant variations in solar activity on time scales of one to three years are filtered by a 24-month Gaussian filter (Hathaway, 2015). Figure 3 shows the sunspot area (dashed-gray line) and sunspot number (black line) series smoothed with a 24-month Gaussian filter for the period 1834–2006.

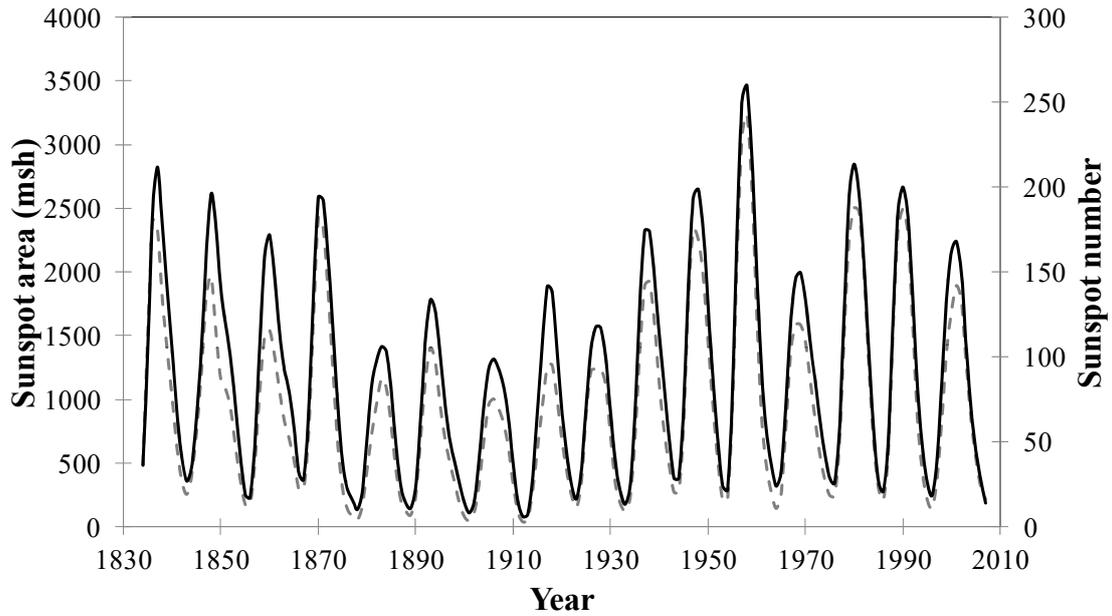

Figure 3. Temporal evolution for annual values of the sunspot-area series normalized (dashed gray line) and sunspot number (black line) series during the period 1834–2006 (Solar Cycles 8–23). Both series have been smoothed with a 24-month Gaussian filter.

In Figure 4, we represent the rise time *versus* maximum amplitude for each solar cycle for sunspot number (top panel) and sunspot area (bottom panel). In the top panel, black squares (open squares) represent the Solar Cycles 8–23 (1–7). In bottom panel, black circles represent the Solar Cycles from 9 to 23, while the open circle corresponds to Solar Cycle 8. This point has been marked thus because its calculation has been slightly different from other points, due to a limitation on the number of data in our series during this period. To establish the value of the minimum of sunspot areas of this cycle and, therefore, calculate the rise time, we have gradually narrowed the range of the Gaussian filter from November 1833. Thus, in the minimum of Solar Cycle 8 (August 1833), the filter is centered in the range $-2a + 5 \leq t \leq +2a - 5$.

Thus, the correlation coefficients obtained for the situations described above are:

- Sunspot number for Solar Cycles 1–23: $r$ = -0.78, *p*-value < 0.001.
- Sunspot number for Solar Cycles 8–23: $r$ = -0.62, *p*-value = 0.010.
- Sunspot area for Solar Cycles 8–23: $r$ = -0.39, *p*-value = 0.135.
- Sunspot area for Solar Cycles 9–23: $r$ = -0.33, *p*-value = 0.227.

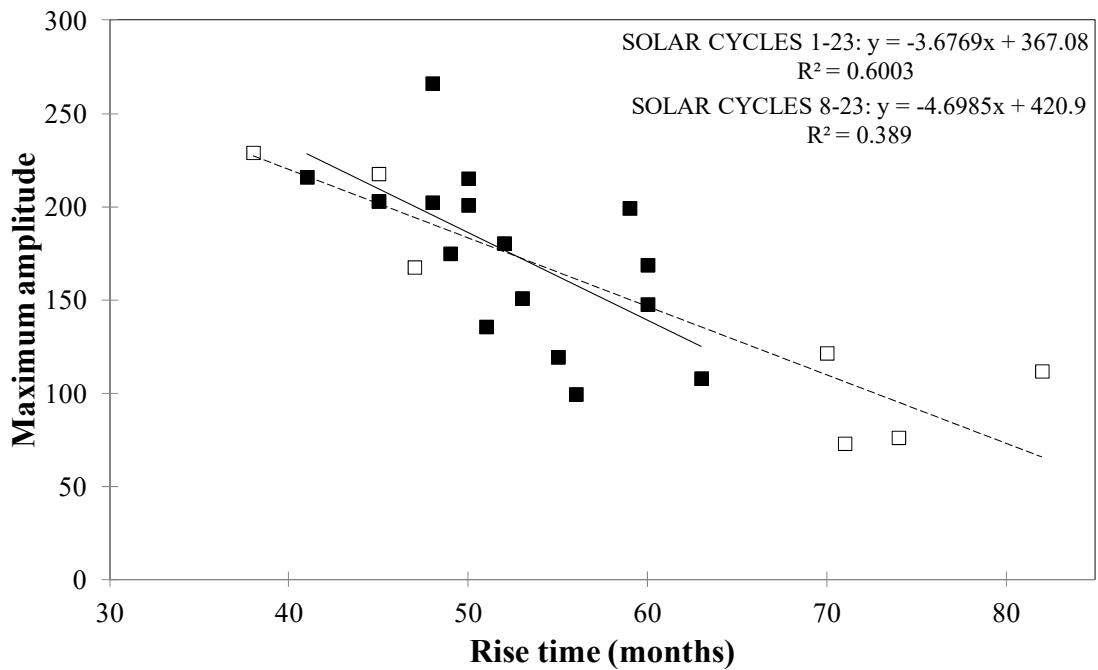

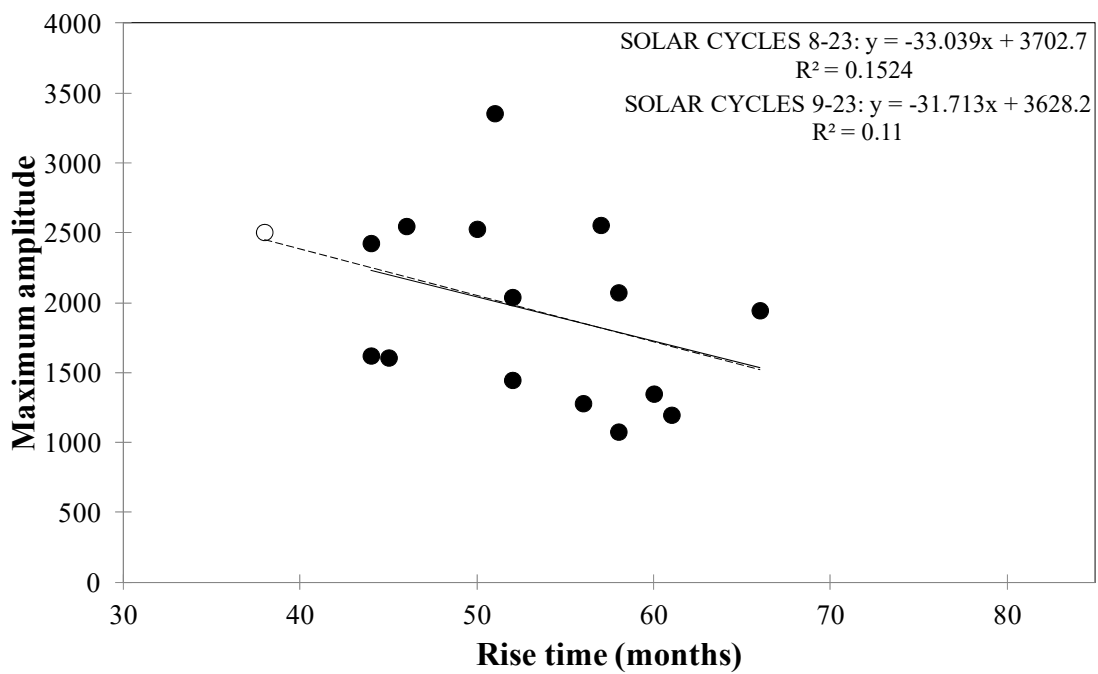

Figure 4. Rise time *versus* maximum amplitude: top panel, sunspot number for Solar Cycles 1 – 7 (open squares) and 8 – 23 (black squares) where dashed line (continuous line) represents the best linear fit for Solar Cycles 1 – 23 (8 – 23) and bottom panel, normalized sunspot-area series for Solar Cycles 8 – 23. The open circle represents Solar Cycle 8. The dashed line (continuous line) represents the best linear fit for Solar Cycles

8 – 23 (9 – 23). The best-fit equations and R-squared coefficients are shown in both panels.

These results show that there is a significant relationship between the rise time and the maximum amplitude of each cycle in the sunspot-number series (although the value obtained for Solar Cycles 8 – 23 is significantly lower than for the whole period 1 – 23) but this fact cannot be seen for the sunspot-area series (Figure 5). Figure 5 shows the family of curves for the sunspot number corresponding to Solar Cycles 1 – 23 (top panel), and the sunspot-area series for Solar Cycles 8 – 23 (bottom panel). It can be seen that while the Waldmeier effect is present in the sunspot number (curves have a similar behavior), it is weaker in sunspot-area series. This result is in agreement with the study of Dikpati, Gilman, and de Toma (2008) where the correlation was calculated using the previous versions of the sunspot areas and the sunspot number. Moreover, we can see in Figure 5 (left panel) that the shape of the solar cycle corresponding to the curve family of the sunspot number is generally asymmetric because the rise time (from minimum to maximum) of the solar cycle is less than the descent time (from maximum to minimum). However, in Figure 5 (right panel), several area curves show an almost Gaussian distribution. Li (1999) asserts this difference is because the reliability of sunspot number is higher than that of the area series because it is easier to measure. Specifically, Pettauer and Brandt (1994) point out that errors in the determination of the sunspot areas are given between 8.5% and 10% even with sophisticated methods.

In addition, we should note that we have also obtained the coefficients of correlation between the rise time and the maximum amplitude of each cycle from the smoothing of the series from 13-month running average and the formula proposed by Meeus (1958). Applying smoothing, the values of correlation coefficients for the area series are lower than those obtained with the Gaussian filter. Thus, from the formula of Meeus and the 13-month running mean, the correlation coefficients are $r$ = -0.28 and $r$ = -0.26, respectively.

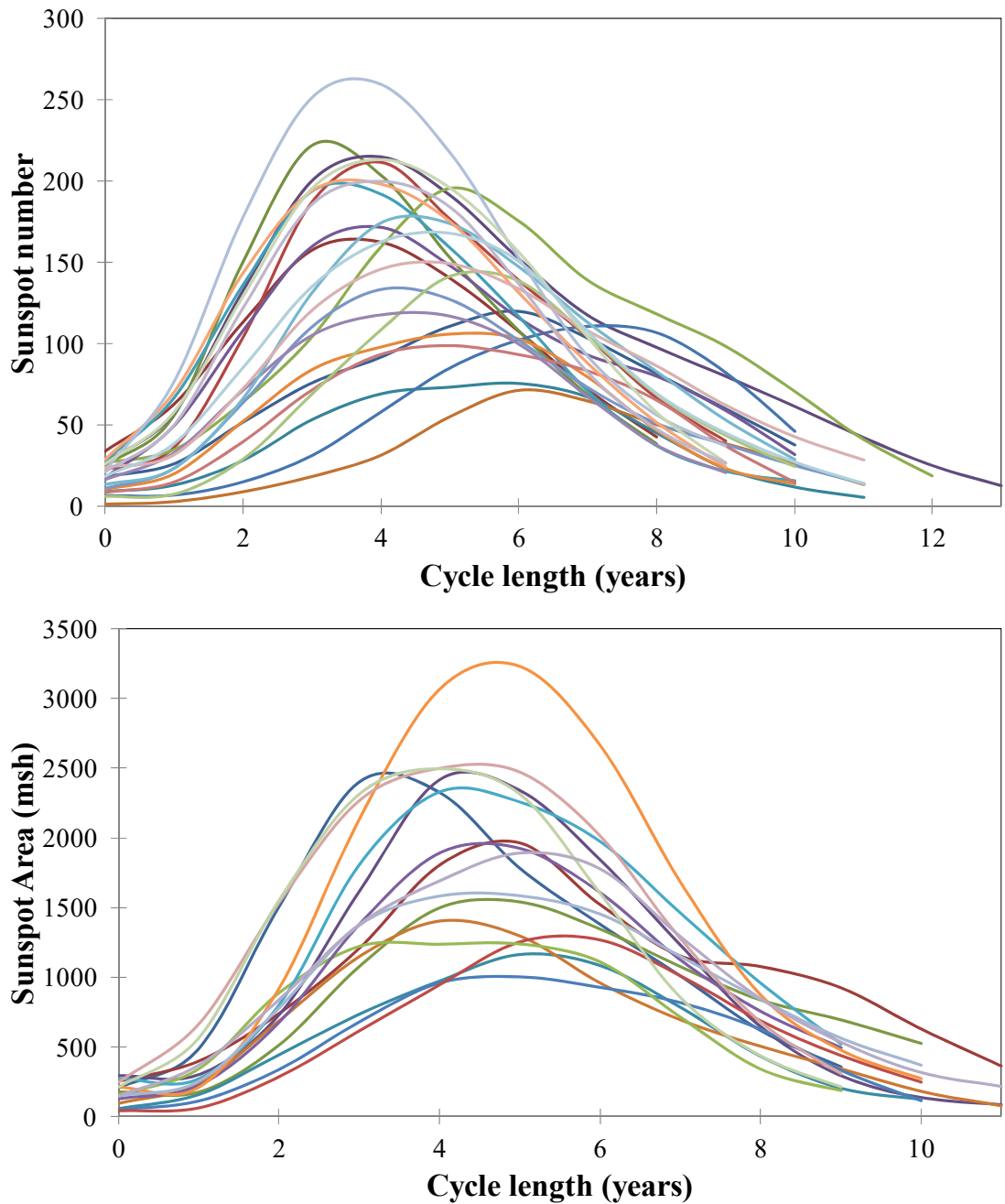

Figure 5. Comparison between curve families: (top panel) sunspot number for Solar Cycles 1 – 23 and (bottom panel) sunspot area series normalized for solar cycle 8 – 23. Both series have been smoothed with a 24-month Gaussian filter.

## 4. Conclusions

A series of bi-weekly records of sunspot areas recorded by Schwabe (1832 – 1853), Carrington (1854 – 1860), and Kew Observatory (1861 – 1868) were compiled by De la Rue, Stewart, and Loewy (1870). In this study, we have constructed a normalized series of monthly sunspot areas by extending the sunspot-area series published by Balmaceda *et al.* (2009) along with the measurements mentioned above. Since there is no temporal overlapping between the two datasets, we have used the new version of the sunspot-number index as link between them, due to the strong correlation between this index and the sunspot-area series. The area series proposed in this article covers almost the last two centuries (1832 – 2008). Thus, this article consists of an update of the study presented by Vaquero, Gallego, and Sánchez-Bajo (2004) using the new versions available for both series.

Furthermore, we have made a spectral analysis in order to find periodicities shorter than the well-known 11-year solar cycle. In addition to the strong signal of the 11-year solar cycle, we have found another periodicity of approximately 300 days around the maximum of Solar Cycles 8 – 9. This 300-day periodicity was also previously detected by other authors (Lean and Brueckner, 1989; Kilcik *et al.*, 2014). In particular, we can note that Kilcik *et al.* (2014) found an approximately 300-day periodicity in the counts of large and well-developed sunspot groups from 1986 to 2013, in agreement with our result. Moreover, other weak signals with periods shorter than 100 days have also been detected in the maximum of the cycle. The results obtained in this study are in agreement with the periodicities found by Vaquero, Gallego, and Sánchez-Bajo (2004).

We have also carried out an analysis of the so-called Waldmeier effect which inversely correlates rise time with the maximum amplitude of a solar cycle. For this purpose, we have utilized the new version of sunspot-number and the sunspot-area series constructed in this study. Thus, we have found a high correlation between the rise time and the maximum amplitude of the solar cycles in the new series of sunspot number ($r$ = -0.78 for Solar Cycles 1 – 23) but not in the number of areas of sunspots ($r$ = -0.39 for Solar Cycles 8 – 23). The values of the correlation coefficients for Solar Cycles 8 – 23 in sunspot number ($r$ = -0.62) and for Solar Cycles 9 – 23 in area series ($r$ = -0.33) are lower than those for the whole datasets used here. As was shown in the above results, correlation coefficients between rise time and sunspot areas are very low compared to sunspot number. This result agrees with the study by Dikpati, Gilman, and de Toma (2008), who concluded that the Waldmeier effect is not present in sunspot-area series.

**Acknowledgements**

Support from the Junta de Extremadura (Research Group Grants GR15137 and GR15146) and from the Ministerio de Economía y Competitividad of the Spanish Government (AYA2014-57556-P) is gratefully acknowledged. We want to thank to the referee for the useful comments that helped to improve this article. Furthermore, we are grateful to the guest editors of this topical issue on Sunspot Number Recalibration (L. Svalgaard and F. Clette) for their useful suggestions.

**Disclosure of Potential Conflicts of Interest**

The authors declare that they have no conflicts of interest.